\documentclass[11pt]{article}
\usepackage{fleqn,y10_style}
\usepackage{url}
\usepackage{graphicx}
\usepackage[figuresright]{rotating}

\def\la{\raise -2.truept\hbox{\rlap{\hbox{$\sim$}}\raise5.truept
\hbox{$<$}\ }}
\def\ga{\raise -4.truept\hbox{\rlap{\hbox{$\sim$}}\raise5.truept
\hbox{$>$}\ }} 
\title{\center CHEMICAL FRACTIONATION AND ABUNDANCES IN CORONAL PLASMA}

\author{\center 
J.J. Drake\address{Smithsonian Astrophysical Observatory, ~Mail Stop 3, 
        ~60 Garden Street, ~Cambridge, ~MA 02138, ~USA}} 

\begin{document}

\maketitle

\centerline{\bf ABSTRACT}

\noindent Much of modern astrophysics is grounded on the observed chemical
compositions of stars and the diffuse plasma that pervades the space
between stars, galaxies and clusters of galaxies.  X-ray and EUV
spectra of the hot plasma in the outer atmospheres of stars have
demonstrated that these environments are subject to chemical
fractionation in which the abundances of elements can be enhanced and
depleted by an order of magnitude or more.  These coronal abundance
anomalies are discussed and some of the physical mechanisms that might
be responsible for producing them are examined.  It is argued that
coronal abundances can provide important new diagnostics on physical
processes at work in solar and stellar coronae.  It seems likely 
that other hot astrophysical plasmas will be subject to similar effects.




\section*{\bf INTRODUCTION} 
\vspace{2mm}

It is difficult to think of an area of astrophysics today in which
plasma chemical composition does not play a key part.  Observations of
the chemical compositions of stars provided the main initial
observational impetus for the development of nuclear
astrophysics. This 
is now used to interpret the observed abundance mixtures and evolution
of stars and interstellar media of galaxies, and by extrapolation, in
intergalactic space.  Along with the theory of nucleosynthesis that
revolutionised astrophysics are underlying assumptions that are still
commonly built into models that use it.  Two simple assumptions are: (i) the
surface chemical composition of a main-sequence star reflects the
initial chemical composition with which that star was
born\footnote{This assumption is of course now modified to allow for
the surface destruction of light elements such as Li, Be and B.}; (ii)
the initial composition of a star is representative of that of the
interstellar medium from which it was formed.

In this context, to my mind coronal abundance anomalies are one of the
more interesting observational developments in the field of stellar
outer atmospheres in recent years.  As element abundances can
provide clues to astrophysical processes such as nucleosynthesis in
stars, coronal abundances should provide us with powerful new
diagnostics of processes occurring in coronal plasmas.  Evidence for
abundance anomalies in the solar corona can be traced back to some of
the first analyses of UV and X-ray spectra to emerge in the 1960s and
1970s (see e.g.\ Meyer 1985 for a review).  These apparent anomalies did not
attract much attention at the time; lingering doubts remained as to
whether or not the observed effects could be attributable to things
other than compositional fractionation, such as a breakdown in the
underlying assumptions of emitting regions comprising optically-thin
plasma in a collision-dominated equilibrium.  It was not until the
1990s that interest in the problem became more widespread.  The
reviews of Meyer (1985) were partly responsible for this, together
with the advent of the X-ray spectrographs on {\it SMM}, {\it Yohkoh}
and of the rocket-borne {\it SERTS} EUV spectrograph, in addition to
solar wind experiments such as {\it Ulysses}.  At the same time,
analyses of {\it Skylab} spectroheliographs were providing convincing
images of coronal abundance variations in the light of lines of
different elements such as Mg and Ne (e.g. Feldman 1992).

Solar anomalies are still largely described in terms of the ``FIP
Effect''---the enhancement of elements with low first ionisation
potentials ($\leq 10$~eV or so; e.g.\ Mg, Si, Fe) relative to those
with high first ionisation potentials ($\geq 10$~eV; e.g.\ H, N, O,
Ne).  Departures from the simple single-step FIP pattern, such as
variations among the high FIP elements, have been observed: indeed,
Schmelz et al.\ (1996) noted {\em ``there is a growing body of
evidence that a simple FIP-based formula is not the whole story for
coronal abundances''.}  As I outline below, element
abundances in the coronae of other stars, and especially the more
active stars, provide modern confirmation of these insightful remarks.

The example of stellar coronae demonstrates that magnetised plasmas
are not necessarily homogeneous in composition, and that significant
chemical separation can occur between regions of different density and
temperature.  As I discuss below, forces are indeed at work to enhance
and deplete elements in coronae: rather than expecting coronae to
share a homogeneous chemical composition with the underlying star, the
issue is more whether or not the corona can mix itself sufficiently
quickly to erase the natural tendency of the plasma to fractionate in
composition.  Whether FIP is really the underlying key parameter in
all cases is not yet clear.  Successful fractionation models must now
have to explain the solar FIP effect as a special case of the
general pattern of abundance anomalies that is emerging in the
coronae of different types of stars.  Similar physics could also be
at work in other hot cosmic plasmas.  Stellar coronae provide a means
for studying these plasma effects and determining in what astrophysical
scenarios some of our most basic assumptions concerning plasma
composition might be incorrect.

\section*{\bf THE FIRST STUDIES OF STELLAR CORONAL ABUNDANCE ANOMALIES: 
EUVE AND ASCA}
\label{s:stellar}
\vspace{2mm}


The last decade ushered in the first detailed stellar coronal
abundance studies, based on the low resolution X-ray CCD pulse height
spectra from the {\it ASCA} satellite and higher resolution {\it EUVE}
reflection grating spectra.  These results have been reviewed recently
by Drake (2002) and are only summarised briefly here. 

EUVE studies showed that stars with
activity levels similar to that of the Sun tend to exhibit a solar-like
FIP effect (e.g.\ $\alpha$~Cen AB, $\epsilon$~Eri, $\xi$~Boo~A), with an
enhancement of low FIP species over high FIP species compared to
photospheric values.  In contrast, active stars appeared considerably 
metal poor with Fe abundances apparently factors of 3-10 below
expectations of their photospheric values.  Schmitt et al.\ (1996), on
finding a coronal Fe abundance for the RS~CVn-type binary CF~Tuc of
5-10 times lower than that of the solar photosphere 
dubbed the phenomenon the {\em Metal Abundance Deficiency (MAD)
Syndrome}.  
At the same time, {\it ASCA} results were also indicating metal
paucity in active stars, from the RS~CVns down to active dMe flare
stars (e.g.\ S.A.~Drake 1996; Singh et al.\ 1999).  This division
between low-activity ``FIP effect stars'' and active MAD stars became
apparent as early as 1995 (Drake et al.\ 1996).  The general picture
of the coronae of active stars being MAD appeared to be confirmed by
analyses of {\it BeppoSAX} LECS spectra of late-type stars in the late
1990s (Pallavicini, Tagliaferri \& Maggio 2000), though there was a
tendency for the {\it BeppoSAX} metallicities to be slightly higher than
those obtained from {\it ASCA} spectra (e.g.\ the case of II~Peg; Covino et
al.\ 2000 vs Mewe et al.\ 1997).  

A body of evidence for {\em apparent} abundance differences between
the plasma in 
large flares and the average quiescent corona of the parent star also
now exists, based largely on {\it ASCA} and {\it BeppoSAX} observations,
but stemming from earlier {\it GINGA} observations of active binaries (e.g.\
Stern et al.\ 1992).  The interpretation in terms of abundances is
likely to be correct (Drake 2002), and provides a tantalising hint that some
large flares are comprised of material evaporated from the
chromosphere, as the ``standard'' flare model (e.g.\ Martens \& Kuin 1989) 
suggests (see also the
contribution of M.~G\"udel in this volume for further discussion of
large stellar flares in this context).

\section*{\bf THE CHANDRA and XMM-NEWTON ERA}
\vspace{2mm}


More recently, {\it Chandra} and {\it XMM-Newton} have been modifying
this picture of coronal abundances anomalies.  The first analyses of
spectra of the active binary system HR~1099 confirmed the low Fe
abundances, but uncovered overabundances relative to Fe of higher FIP
elements and especially large enhancements of Ne (Brinkman et al.\
2001; Drake et al.\ 2001).  Drake et al.\ (2001) showed that evidence
for the high Ne abundances was indeed present in earlier {\it ASCA}
analyses of active stars in general but had been ignored.  Brinkman et
al.\ (2001) suggested the anomalies in HR~1099 followed an {\em
inverse} FIP effect in which high FIP elements were enhanced relative
to low FIP elements.  Audard and co-workers (see contribution in this
volume) suggest, based on a more extensive analysis of XMM-Newton
observations of stars of different activity level, that there is a
continuum of behaviour from the FIP effect in relatively inactive
stars like the Sun to the inverse FIP effect in the most active
stars.  
However, the patterns do not all appear to be so simple to characterise:
significant deviations from a FIP-based formula are present in the
data both for HR~1099 (Brinkman et al.\ 2001; Drake et al.\ 2001), and
from more recent analyses of the active binaries II~Peg and AR~Lac
(Huenemoerder et al.\ 2002,2003).  In particular, the abundances
derived in the latter study of AR~Lac indicate higher values for Al
(FIP=5.99~eV) and Ca (FIP=6.11~eV) than for the 
elements Mg, Si, and Fe that all have a 
slightly higher FIP, opposite to the pattern expected for an
inverse FIP effect.  In HR~1099, Fe appears anomalously low compared
to elements Si amd Mg with similar FIP, suggesting that the
fractionation might not be completely mass-independent as is believed
to be the case for the Sun.

The pattern of abundance anomalies across different star types and
activity levels now appears much more rich and diverse than we
realised.  An important question is whether or not the anomalies are
predictable based on observed stellar indices, or whether they depend
so critically on coronal characteristics that in integrated starlight
they can visibly change on short timescales on any given star, even
outside of flares.  The solar case does not provide a reliable guide
here: observations indicate strong variations in abundances in
different regions of the solar corona (e.g.\ Meyer 1996), and
abundances that are seen to change on timescales of days in some
coronal structures, such as emerging active regions (e.g.\ Sheely 1995).
However, it is important to note that {\em there are no full-disk
measurements of solar coronal abundances}, so it is difficult to
estimate how the variations in abundances in the solar corona might
translate to the integrated disk.  In the stellar case, there are very
few, if any, repeat observations of the same star made with the same
instrument.  Cross-comparison between different instruments can be
difficult owing to the different uncertainties inherent in the
analyses (e.g.\ Drake 2002).

That integrated disk abundances are in fact quite predictable is
supported by a recent comparative study of the active binaries Algol
and HR~1099 (Drake 2003).  The dominant components of these systems at
X-ray wavelengths (G8~III and K1~IV in Algol and HR1099, respectively)
have very similar effective temperatures (4500~K, 4750~K), radii
($3.5R_\odot$, $3.7R_\odot$), masses ($0.8M_\odot$, $1.0M_\odot$) and
rotation periods (2.87d, 2.84d); it might then be expected that they
also share similar magnetic activity characteristics.  The {\it
Chandra} LETG spectra of these systems (Figure~1)
are indeed remarkably similar over a broad range of charge states of
Fe as well as in lines of and H- and He-like ions of O, Ne, Mg and Si.
Both systems share a similar depletion of Fe by about a factor of 3,
and strong enhancements of Ne.

\begin{figure}
\includegraphics[width=2.5in,angle=270]{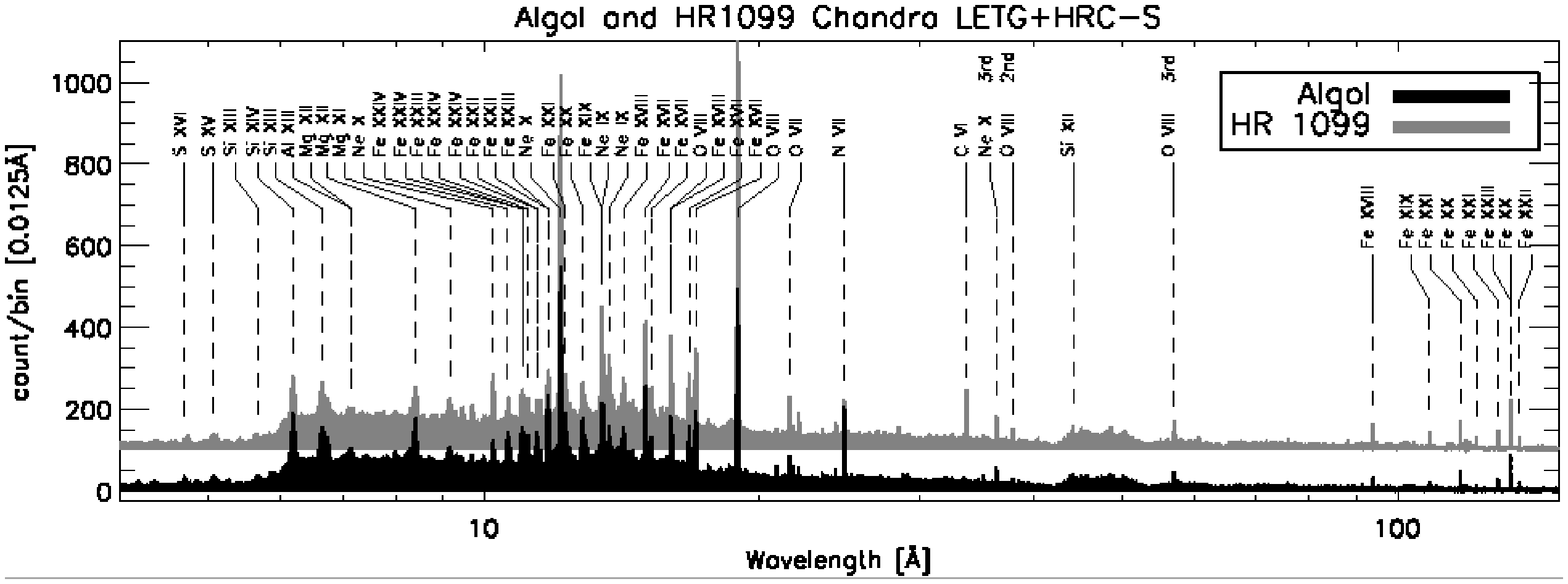}
{\sf Fig.~1.~Illustration of the remarkable similarity in the combined
positive and 
negative order LETG+HRC-S spectra of
Algol and HR1099 binned at 0.0125~\AA\ intervals.}
\label{f:algol_hr1099}
\vspace{-2mm}
\end{figure}

\section*{\bf ``NEW'' TECHNIQUES FOR HIGH RESOLUTION:
TEMPERATURE-INSENSITIVE LINE RATIOS} 
\vspace{2mm}


Once spectral lines can be easily resolved, as exemplified by the {\it
Chandra} grating spectra,\footnote{This is strictly only partially true for
{\it XMM-Newton} because a significant fraction of the flux in 
spectral lines observed with RGS remains unresolved in the broad wings
of the instrumental profile.} abundance analysis becomes
at once both more obvious and more complex: while lines due to
different elements can be discernible, making good use of the
diagnostics usually requires a much more involved analysis than the
relatively crude two-temperature parameter estimation methods used for
low-resolution spectra.  The source plasma is, instead, typically
multi-thermal (e.g.\ Drake 1996), and the observed spectral lines
containing the abundance information are formed over temperature
ranges that can cover substantial variations in source plasma emission
measure.  Consequently, in the general case, abundances can only be
derived if the source emission measure distribution is first
determined (which is, in any case, notoriously difficult to determine
if true uncertainties are to be accounted for---see Kashyap \& Drake
1998 for a recent discussion).  Such an analysis can be very
time-consuming owing to the large number of lines that must be
included from the different charge states present in the multi-thermal
source.

One simplifying approach to analyses of solar EUV and X-ray spectra
uses the ratios of spectral lines formed at similar temperatures.  In
the approximation that the ratio of the emissivities of two lines from
different elements is constant over the temperature of formation of
the lines involved, the ratio of the observed line fluxes yields
directly the element abundance ratio.  Such an approach can greatly
simplify abundance analysis because the underlying emission measure
distribution need not be know.  However, as discussed by Drake et al.\
(2003), the ratios that have been used for solar X-ray spectra in the
past are not so well suited to the stellar case because emissivity
ratios depart significantly from a constant value over the range of
temperatures seen in stellar X-ray spectra.  

\begin{figure}

\begin{minipage}{75mm}

\includegraphics[width=70mm]{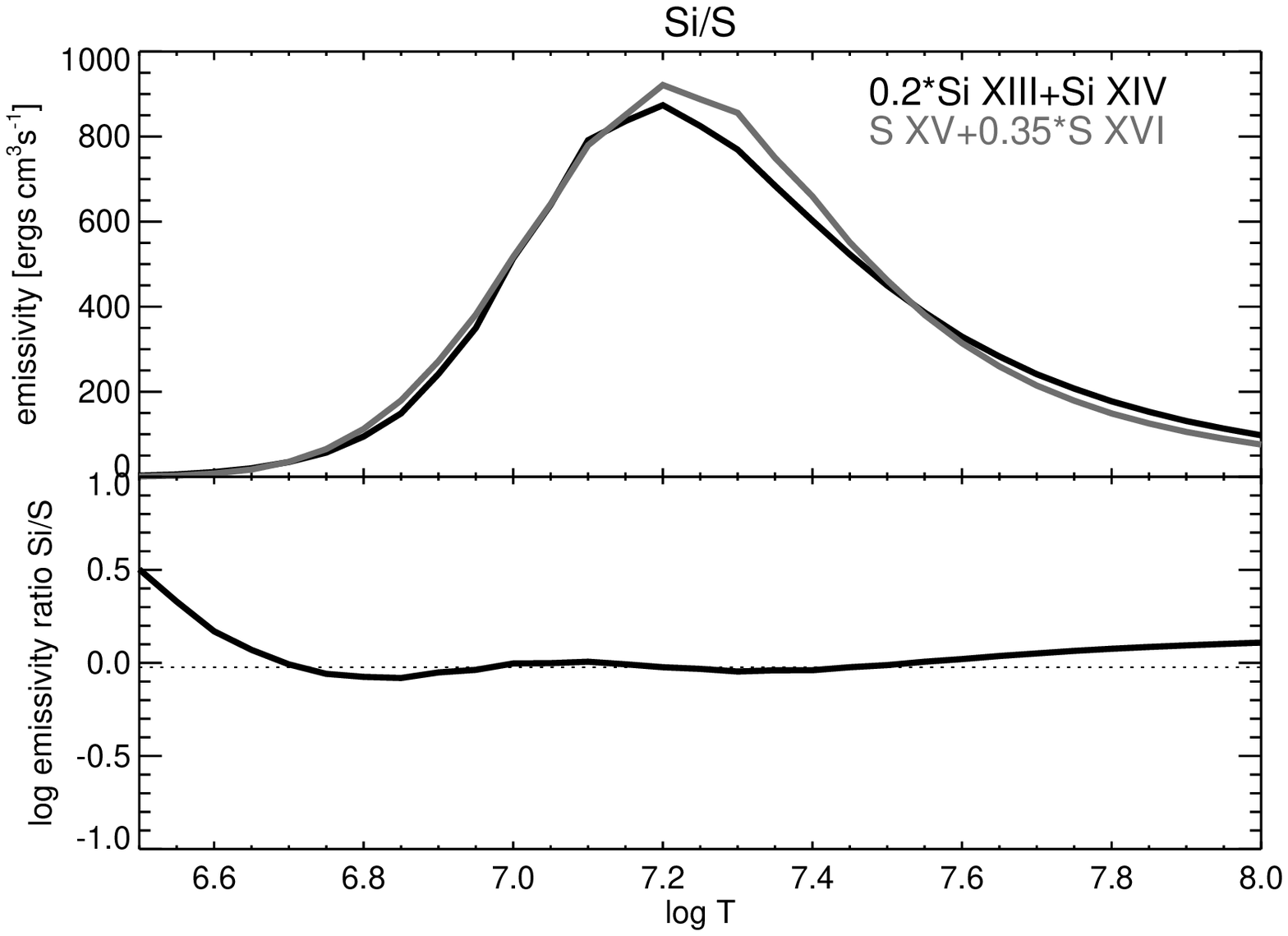}
{\sf Fig.~2.~Emissivities and their ratio vs temperature for a mixture of 
the Si and S H-like and He-like resonance lines.  The ratio forms a 
temperature-insensitive abundance diagnostic.}
\end{minipage}

\hfil\hspace{\fill}

\hspace{76mm}
\begin{minipage}{100mm}
\vspace{-80mm}
{\sf Table 1. Abundance ratios for AU Mic
based on temperature-insensitive H- and He-like line ratios, 
together with prominent lines of Fe (Drake et al.\ 2003).  Quoted
uncertainties are from Poisson statistics only; true uncertainties
will be larger by typically 0.1~dex.}
{\scriptsize
\begin{tabular}{lc}
\hline\hline
Diagnostic & Abundance Ratio$^a$ \\ \hline
 N VII/(O VII+0.35*O VIII) &$[$N/O$]_{\odot}=0.10\pm   0.12$\\
 O VII/(Ne IX+0.15*Ne X)   &$[$O/Ne$]_{\odot}=-0.29\pm 0.10$\\
 (0.07*Ne IX+Ne X)/(Mg XI+0.17*Mg XII)  &$[$Ne/Mg$]_{\odot}=0.61\pm 0.06$\\ 
 (0.35*Ne IX+Ne X)/Fe XVII            &$[$Ne/Fe$]_{\odot}=0.77\pm 0.04 $\\ 
(0.15*Mg XI+Mg XII)/(Si XIII+0.2*Si XIV) & $[$Mg/Si$]_{\odot}=-0.29\pm 0.06$\\ 
 Mg XI/(0.5*Fe XVII+Fe XXI+2.0*Fe XVIII)  & $[$Mg/Fe$]_{\odot}=0.03\pm 0.10$\\
(0.2*Si XIII+Si XIV)/(S XV+0.35*S XVI)   & $[$Si/S$]_{\odot}=-0.07\pm 0.17$\\ 
(0.4*Si XIII+ Si XIV)/ Cont.\ at $5.85 \pm 0.5$~\AA
&$[$Si/H$]_{\odot}= -0.43 \pm 0.10$\\ 
\hline
\end{tabular}
} 
{\sf ~~~$^a${\footnotesize Ratios are relative to solar values and
are  expressed in the standard logarithmic bracket notation.}\\
}

\end{minipage}

\vspace{-2mm}
\end{figure}

Due to the different temperature dependencies of spectral lines
arising from different elements, it is obviously not possible to
construct indices that are perfectly insensitive to temperature.
However, within temperature intervals where the coronal emission
measure is expected to be significant for active stars ($\sim
10^6$-$10^7$~K), Drake et al.\ (2003) have found that it is possible
to find combinations of ratios of lines of different elements that are
constant within reasonable uncertainties ($\la 50$\%\ or so) for their
respective temperature ranges of formation.  These ratios are
constructed from, typically, two lines of each element, mixed such
that the summed emissivity profiles in the numerator and denominator
follow each other much more closely than ratios formed from single
lines.  An example for Si and S based on their He- and H-like
resonance lines is illustrated in Figure~2.  

This line
ratio technique has been applied to the {\it Chandra} HETG spectrum of
the active M0 dwarf AU Mic; the diagnostics and results are listed in
Table~1.  These results demonstrate immediately that
AU~Mic does share some similarities in coronal composition with the
RS~CVn-type binaries: enhanced Ne and depleted Fe relative to elements
such as O, Mg and Si, for an assumed photospheric composition similar
to that of the solar photosphere.  The Si/S ratio for which the
emissivity curves are illustrated in Figure~2 shows that
both have essentially the same abundance, despite a difference in FIP
(Si 8.15 vs S 10.36~eV).  Uncertainties in these abundances tend to be
dominated by the small residual temperature dependence of the
emissivity ratios rather than by Poisson statistics in these bright
lines, but still amount to less 50~\%\ in general (Drake et al.\
2003).


\section*{\bf PROCESSES TENDING TO FRACTIONATE AND HOMOGENISE STELLAR
OUTER ATMOSPHERES}  
\vspace{2mm}


\noindent{\bf Fractionation Forces}

The solar corona is an inhomogeneous, dynamic, magnetised plasma.  The
source of coronal plasma is the underlying, weakly ionised
chromosphere/photosphere.  In the passage from the photosphere to the
corona, and in the corona itself, forces are at play that do not act
equally on all particles.  The most relevant quantities dictating the
different forces on a particle are its charge and mass.  These forces
will tend to segregate the plasma unless there are other processes at
work to homogenise it.  The principal forces at work are listed
below. 
\begin{equation} 
\;\;\;\;\;\;\;\;\;\;\; {\rm Gravitational\;\; settling} 
\;\;\;\;\;\;\;\;\;\;\;\;\;\;\;\;\;\;\;\;\;\;\;\;\;\;
 m_ig
\end{equation}
\begin{equation} 
\;\;\;\;\;\;\;\;\;\;\; {\rm Thermal
\;\;diffusion}\;\;\;\;\;\;\;\;\;\;\;\;\;\;\;\; 
- \frac{6}{5}\frac{\nu_{ie}\mu_{ie}}{kT}\frac{\rho_i}{\rho_e}\kappa _e
T^\frac{5}{2} \; \frac{dT}{ds}
\end{equation}
\begin{equation} 
\;\;\;\;\;\;\;\;\;\;\; {\rm Ambipolar\;\; electric\;\; field}
\;\;\;\;\;\;\;\;\;\;\;\;\;\;\;\;\;\;\;\;\;\; Ez_i
\end{equation}
And when flows are involved, 
\begin{equation} 
\;\;\;\;\;\;\;\;\;\;\; {\rm Frictional\;\;drag}
\;\;\;\;\;\;\;\;\;\;\;\;\;\;\;\;\;\;\;\;\;\;\;\;\;\;\;\;\;\;\;\;\;\;
m_i\nu_i\delta u_i
\end{equation}

In these equations, subscripts $e$ and $i$ denote electrons and ions,
respectively, $m$ is the particle mass, $\mu$ is the reduced mass, $g$
is the component of gravity parallel to the magnetic field that
defines the coronal structure in question (such as a loop), $s$ is the
distance along the magnetic field, $\rho$ is the species density,
$\nu$ is the collision frequency, $u$ is the flow speed, $T$ is the
temperature, $z$ is the number of ionized electrons, $E$ is the
polarization electric field, $\kappa$ is the coefficient of thermal
conductivity and $k$ is Boltzmann's constant.

\vspace{2mm}
\noindent{\underline{Gravitational Settling}~~}

The pressure scale height for a given species is proportional to the
particle mass.  In order to prevent mass-dependent
gravitational settling of metals, the corona must either be mixed,
e.g.\ by turbulence, undergo bulk flow (including syphon flows) with a
velocity larger than the gravitational settling velocity, or else
gravity must be balanced with an outward-directed force. 

\vspace{2mm}
\noindent{\underline{Thermal Diffusion}~~}

Also known as the ``thermal force'',
thermal diffusion refers to the tendency in regions of a plasma with a
temperature gradient for heavier ions to diffuse toward higher
temperature.  This is essentially a kinetic effect caused by the effective
non-Maxwellian velocity distribution of electrons in the region of the
temperature gradient.  Thermal diffusion was initially postulated as a
possible explanation for the apparent excess of some metals in the corona
found in early spectroscopic studies (e.g.\ Delache 1967; Nakada 1969;
Tworkowski 1976).

\begin{figure}
\includegraphics[width=1.6in,angle=90]{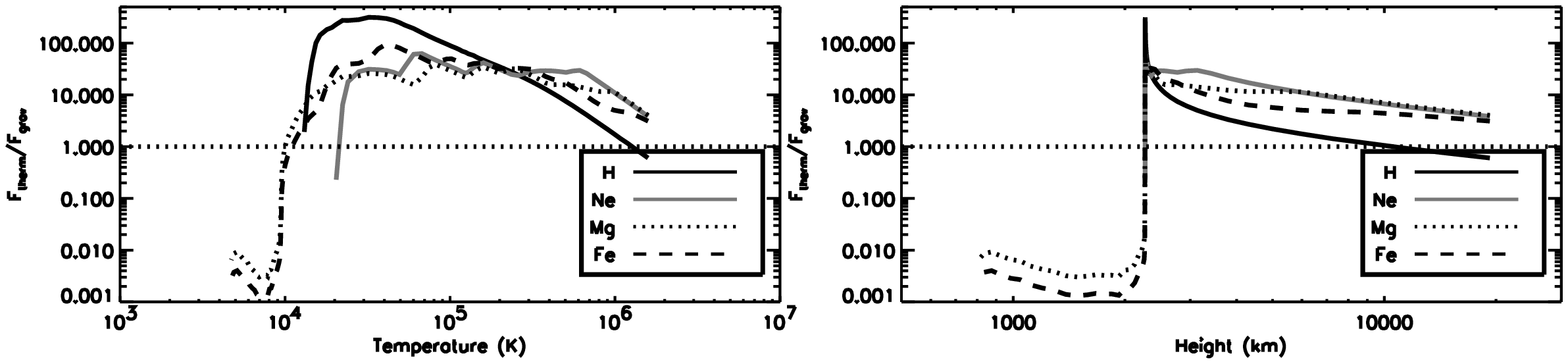}
{\sf Fig.~3.~Ratio of thermal to gravitational forces for H, Ne, Mg and
Fe as a function of plasma temperature and altitude in the VALC solar model.}
\vspace{-2mm}
\end{figure}

However, thermal diffusion acting alone against gravity tends to
enhance all metal abundances in the corona and does not easily produce a FIP
bias.  This can be seen by calculating the ratio of thermal to
gravitational forces for H and different metals.  I illustrate this
ratio for H, Ne, Mg and Fe in Figure~3 for the VAL~C
model solar atmosphere (Vernazza et al.\ 1981).  It can be seen that
gravity dominates until the onset of the strong temperature gradient
at the base of the transition region.  At this point, thermal
diffusion is two orders of magnitude stronger than gravitational pull
for H, and rises to more than an order of magnitude over gravity for metals.
Based on these calculations, it appears unlikely that metals can be 
enhanced by thermal diffusion until the upper transition region at
temperatures of a few $10^5$~K, because at lower altitudes H is more
strongly pulled into the corona---this would lead instead to a
metal-poor plasma.  At coronal temperatures, the thermal force does
tend to select metals; however, there is another important force
that acts on ions to further complicate the fractionation process.

\vspace{2mm}
\noindent{\underline{Ambipolar Diffusion}~~}

The electric field, $E$, arises in response to the tendency of the
plasma to differentially stratify through thermal diffusion and
gravitational settling.  To first order, the ambipolar field is due to
the spatial distribution of protons and electrons.  An obvious example
of the significance of the ambipolar field is in holding down the
electrons in the corona.  The scale height for electrons is $m_e/m_p$
times that for protons, but in large loops this tendency to separate
is counterbalanced by the electric field thus generated.  At
the top of the corona, the ambipolar electric field is therefore
directed outward.  At lower altitudes the situation is more
complicated because of the balance between the thermal force and
gravity.  In strong temperature gradients where the net force on
protons is upward, then the expectation is for an electric field
directed downward.  These expectations are borne out by detailed
calculations for the solar outer atmosphere by Fontenla (2002), and  
for multi-species loop models  (Lenz 1999).

The net influence of the ambipolar field on a species $i$ depends on
the charge-to-mass ratio, $q_i/m_i$.  Since the electric field arises
to counterbalance other forces that tend to separate protons and
electrons, the direction of the electric force on metals relative to H
is given by their relative charge-to-mass ratios $q_im_H/m_iq_H$.  All
metals in all ionisation stages have a lower $q/m$ than ionised H, so
their acceleration in the electric field is smaller.  There are two
important limits to the charge-to-mass ratio, $q_i/m_i$: (i) near the
base of the transition region where low FIP light elements are likely
to be twice ionised, but high FIP species only once ionised, $q_i/m_i$
is higher for low FIP elements by a factor of $\sim 2$; (ii) at higher
temperatures, $q_i/m_i$ is more similar for both high and low FIP
elements, and tends to $\sim 2$ when elements are fully ionised.  At
the top of the corona, where the ambipolar field is directed outward,
all metals will therefore tend to sink relative to H.  Note that this
tendency is opposite that of thermal diffusion
(Figure~3).  In the transition region, where the
ambipolar field is expected to be directed down toward the
chromosphere, metals are drawn down to a lesser extent than H by the
electric field (but more by gravity), and the net acceleration
relative to H then depends on the balance between thermal, electric
and gravitational forces.


\vspace{2mm}
\noindent{\underline{Frictional Drag}~~}

The different forces described above can induce migrations of
different species relative to each other.  Such flows are subject to
frictional forces that act, in general, to reduce relative flow.  In a
plasma dominated by hydrogen, friction for a given species arises
through collisions with hydrogen atoms or protons.  The frictional
force depends on the collision frequency.  Collision rates are much
larger between protons and ions than that between protons and
neutrals, or between hydrogen atoms and other neutrals or ions.  These
differences will be most crucial in the chromosphere and lower
transition region where neutrals and ions co-exist with ionised
hydrogen; upward drift of protons, as a result of mass-loss in a wind
for example, will more easily carry ions with it than neutrals.  In
such a scenario, elements with FIP$>13.6$~eV can be left behind.  This
mechanism provides a basis for some models seeking to explain the FIP
effect in the solar corona and wind (e.g.\ Marsch et al.\ 1995; Peter
1998) though McKenzie (2000) has shown that such 1-D models rely on spurious
boundary conditions.  Schwadron et al.\ (1999) present a detailed
discussion of frictional effects and show that moderate flows of only
a few km/s can be effective in eliminating compositional
fractionation.  These models included the ambipolar field due to
hydrogen, though the thermal force was neglected.

\vspace{2mm}
\noindent{\bf  Anti-Fractionation Processes}

High resolution {\it TRACE} image sequences of the solar corona
discussed elsewhere in this volume serve as a dramatic demonstration
that coronae are dynamic systems characterised by (possibly
ubiquitous) flows, and perhaps also by significant turbulence. Flows
and turbulence are homogenising agents, acting against the processes
discussed above that tend to chemically separate the plasma.

\vspace{2mm}
\noindent{\underline{Turbulence}~~} 

In the case of turbulence, on scales smaller than the particle
mean-free path (``microturbulence'') the turbulent velocity would need
to be comparable to the thermal velocity to induce sufficient mixing.
As noted by Schwadron et al.\ (1999), such turbulence can probably be
ruled out.  However, macroturbulent mixing on scales much larger than
the particle mean-free path only need occur with velocities comparable
to that of gravitational settling or ambipolar and thermal diffusion
in order to prevent it.  

\vspace{2mm}
\noindent{\underline{Flows}~~}

Stellar coronae are in a state of flow because they are the source of 
mass loss in a wind---the slow wind in the case of closed,
X-ray-bright magnetic structures on the Sun. 
As discussed above, flows could act to fractionate the plasma through
differential frictional forces.  However, once flows are significantly
more rapid than migration caused by forces acting opposite to the
flow---such as gravity in the case of upward flow---then all particles
are carried along and the composition remains the same as that at the
base of the flow (again, see the insightful discussion of Schwadron et
al. 1999).  The gravitational settling time for typical loops with
densities of $10^9$-$10^{10}$~cm$^{-3}$ in the solar corona is of
order a day or so.  Schwadron et al.\ (1999) estimate that flow speeds
as low as $\sim 100$~ms$^{-1}$ can be sufficient to prevent
gravitational or mass-dependent settling.  Interestingly for the case of
enhanced Ne and other high FIP elements in the coronae of active
stars, Schwadron et al.\ (1999) find that similar effects in their
model can be produced by moderate syphon flows.

\section*{\bf COMMENTS ON SOME FRACTIONATION MODELS}
\vspace{2mm}


The advent of high resolution X-ray spectra of stellar coronae that
allow the measurement of coronal abundances with greater accuracy and
reliability than earlier generation instruments provides new
challenges to fractionation models.  Stellar observations show that
the solar case is evidently only one particular example in the
apparent continuum of abundance anomalies.  Models proposed to date
have been aimed at explaining the solar FIP Effect and enhancement of
low FIP species.  I refer to the nice summary of some of these models
by Henoux (1995) for more detailed discussion.  A successful model now
has to be able to reproduce this range of anomalies.  A full
examination of the different models that have been proposed is beyond
the page limit here; I comment instead on one or two models concerning
fractionation that have appeared in the recent literature.


\vspace{2mm}
\noindent{\underline{Neutral-ion separation in a magnetic field}~~} {\it e.g.\
Vauclair (1996); Steinitz \& Kunoff (1999)}\\ 
Magnetic field-based separation acting in the chromosphere relies on
ions being tied to the field but neutrals not.  Vauclair (1996)
proposed that rising magnetic field preferentially carried ions with
it into the corona.  Steinitz \& Kunoff (1999) invoked the diamagnetic
effect, or ``magnetic mirror'', whereby field lines diverging after
emerging from the photosphere preferentially accelerate ionised
species upward.  Both models have problems if high FIP species are to
be preferentially brought into the corona---the magnetic field instead
needs to behave in the opposite fashion to retard ions.  Doubtless it
is possible to conceive a gradual change in magnetic field behaviour
from solar-like activity levels to the most active stars that might achieve
this, though it does not seem straightforward.

\vspace{2mm}
\noindent{\underline{Multi-species loop models}~~} {\it Lenz (1999)}\\
A nice discussion of the forces acting on different species in coronal
loops has been presented by Lenz (1999), who investigated
self-consistent loop models subject to different heating conditions.
The abundance profiles of different elements in these quasi-static
models was found to be sensitive to loop parameters, such as length,
base pressure and heating rate, illustrating how plasma can
fractionate in the absence of mixing processes.  However, these
models are static and the abundance profiles obtained can be
substantially different in the presence of flows.

\vspace{2mm}
\noindent{\underline{Flows and wave heating}} {\it Schwadron et al.\ (1999)}\\
Schwadron et al.\ (1999) studied the impact of
flows and the resulting ambipolar field on species populations through
a coronal loop.  These effects alone do not produce a low FIP bias,
though syphon flows could raise coronal abundances of high FIP
elements---reminiscent of the situation in active stars.  However, the
same model could obtain enhancements of low FIP elements by applying
MHD wave heating: ions are heated with respect to neutrals and so low
FIP species can attain greater scale heights.  Thus---by
accident---this type of model is able to reproduce some of the aspects of
coronal abundance anomalies in both low and high activity stars.
While this warrants further study, it remains questionable whether
coronae are too dynamic to be fractionated by these means, and, as
argued by McKenzie (2000), whether time-dependent or spatially
inhomogeneous processes are responsible.

\section*{\bf CONCLUSIONS} 
\vspace{2mm}


Coronal plasma under the influence of gravity and/or with a
temperature gradient will tend to chemically fractionate unless mixed
or subject to flows.  Magnetic fields might also play crucial roles in
separating neutral and ionised species.  Abundance peculiarities are
then to be expected in coronal plasma, hopefully providing us 
with new diagnostics.  
Unfortunately, the interplay of the different forces and processes
involved is complex, and the situation is likely complicated by the
effects of flows and turbulence, and possibly by dynamic magnetic
fields.  Reading the observed chemical composition will require
greater understanding of the controlling processes.  Successful models
describing coronal abundances are now need to satisfy the range of
abundance anomalies being uncovered by stellar
observations---enhancements of high FIP elements and apparent
depletions of low FIP elements relative to photospheric
abundances---in addition to the solar FIP effect.

However, despite the complicated physics and current lack of understanding of
abundance anomalies, we are already in a position to make some useful
statements.

\vspace{2mm}
\noindent{\underline{Coronae are not hydrostatically stratified}~~}

The scale
height for species in the solar corona is $\sim 30,000T_6/m$~km where
$T_6$ is the species temperature in units of $10^6$~K and $m$ is in
amu.  For protons at a typical temperature of $2\times10^6$~K, the scale
height is then 60,000~km.  As noted earlier, the scale height for Fe
is 1/56 that of H, or about 1000~km for the same temperature, and so large
long-lived coronal loops would therefore be essentially empty of Fe if
gravitational settling were not counteracted by opposing forces or by
mixing.  An obvious conclusion that can then be drawn from the
observation of iron at super-photospheric or photospheric abundance in
large coronal loops is that the solar corona is not generally in
hydrostatic equilibrium.  

\vspace{2mm}
\noindent {\underline{Mixing processes must be at work in low gravity active
stars and in rapidly rotating stars}~~~} 

Typical giant stars can have
surface gravities 100 times less than that of the Sun.  Such low
effective gravities can also be achieved on less evolved stars that
are very rapidly rotating.  These active stars tend to have coronae
hotter than that of the Sun by an order of magnitude, with typical
temperatures of $10^7$~K.  Upward thermal forces in these coronae will be
much stronger than for the Sun, while the downward gravitational force
can be much weaker.  Since we do not see active coronae filled with metals
(rather, they appear depleted in metals), their coronae must be 
significantly mixed by turbulence or homogenised by flows.  The models
of Lenz (1999) also support this conclusion for the solar corona since
quite different abundance patterns that are not observed 
were produced for different loop lengths and base pressures.  This
indicates that something else is driving the observed composition.

\vspace{2mm}
\noindent{\underline{Magnetic field fractionation mechanisms}~~}

If the
magnetic field is indeed responsible for fractionation in the
chromosphere, then coronal abundances provide will insights into
changes in magnetic field topology as a function of stellar activity
level.

\vspace{2mm}
\noindent{\underline {Limits on flow velocities}~~}

If syphon-type flows are responsible for the high FIP biases seen in
active stars, as hinted by the models of Schwadron et al.\ (1999) for
solar loops, then we can place limits of a few km/s on the velocities
of such flows since significantly larger velocities should erase the
abundance pattern.

\vspace{2mm}
\noindent{\underline{Wave heating and low FIP enhancements}~~}

 The model of
Schwadron et al.\ (1999) invokes preferential MHD wave heating of
ionised species in the chromospheric region of partial ionisation to
explain enhanced low FIP elements.  It is unlikely that such heating
mechanisms can be significant over the full range of activity levels
observed for late-type stars (e.g.\ the most active stars have X-ray
luminosities $10^4$ times higher than the Sun).  We might then
expect a qualitative change in observed anomalies as such wave heating
becomes insignificant, much as is observed.

These points provide only very simple examples of the insights
abundances might provide for understanding the physics of coronal
plasma.  As noted in the introduction, similar fractionation processes
are likely to be at work in other hot cosmic plasmas.  Stellar coronae
provide laboratories for studying these interesting and important
effects.
 
\section*{\bf ACKNOWLEDGEMENTS}
\vspace{2mm}


I am happy to acknowledge insightful conversations with my tolerant 
collaborators J.M.~Laming and V.~Kashyap during the course of this
work.  I am grateful to L.~Lin, who contributed to the analysis
AU~Mic, and thank the editor, Louise Harra, for her patience.
JJD was supported by NASA contract NAS8-39073 to the {\em
Chandra X-ray Center}.

\end{document}